\documentclass[prc,showpacs,aps,twocolumn]{revtex4}
\usepackage{graphicx}

\begin{document}
\title {RPA vs. exact shell-model correlation energies}
\author{Ionel Stetcu}
\altaffiliation{On leave from National Institute for Physics and
Nuclear Engineering -- Horia Hulubei, Bucharest, Romania.}
\author{Calvin W.~Johnson}
\altaffiliation{Current address: Physics Department, 
San Diego State University,
5500 Campanile Drive, San Diego CA 92182-1233}
\affiliation{
Department of Physics and Astronomy,
Louisiana State University,
Baton Rouge, LA 70803-4001
}

\begin{abstract}
 The random phase approximation (RPA) builds in correlations left
out by mean-field theory. In full $0\hbar\omega$ shell-model spaces we
calculate the Hartree-Fock + RPA binding energy, and compare it
to exact diagonalization. We find that in general HF+RPA gives a 
very good approximation to the ``exact'' ground state energy. 
In those cases where RPA is less satisfactory, however, there is no 
obvious correlation with properties of the HF state, such as 
deformation or overlap with the exact ground state wavefunction. 
\end{abstract}


\maketitle

\section{introduction}
\label{sec:intro}

One objective of nuclear theory, not reached yet, is an accurate
method to compute the binding energy of nuclides globally--that
is, from helium to the superheavies beyond uranium,  and to 
the driplines relevant to
nucleosynthesis. The most successful and widely referenced
approaches are semiclassical, such as the finite-range droplet model \cite{MN95} 
and the extended Thomas-Fermi plus Strutinski-integral model \cite{AP95}.  
Exact, fully quantum mechanical \textit{ab initio}
microscopic calculations starting from experimental NN scattering
have been successful \cite{benchmark,Aeq8nuclei,Aeq12nuclei} 
but are computationally too
demanding for all but the lightest nuclei.

What about less ambitious but still microscopic approaches? 
The method of choice for a global description 
has been mean-field models such as Hartree-Fock 
(HF) with a phenomenological interaction (e.g. Skyrme) \cite{HFmass}.
For computational simplicity Hartree-Fock is easiest to apply to 
spherical nuclei, although  advances in computers allow one to 
treat deformed HF as well; in that case the deformed HF state breaks 
rotational symmetry. Other symmetries that are broken by Hartree-Fock 
include translational invariance and (for Hartree-Fock-Bogoliubov) 
particle-number conservation. Mean-field theory thus leaves out both 
``kinematic'' correlations, which restore broken symmetries and which 
must be corrected for approximately \cite{HFmass}, and ``dynamical''
correlations, which arise from the residual two-body interaction. 

 The interacting shell
model (SM), which diagonalizes an effective Hamiltonian in a basis
of shell-model configurations, includes correlations, respects
rotational invariance and other symmetries, and yields accurate binding
energies, but at a computational price.  Shell model calculations are currently
limited to about $10^{7-8}$ basis states, which restricts one to
light to medium nuclei ($A < 50$) for full $0\hbar\omega$ calculations.
Furthermore, there is no
global shell model interaction; instead one must derive or fit an
effective interaction for each regime such as the $sd$-shell (with
valence nucleons between magic numbers of 8 and 20) and the
$pf$-shell (between magic numbers of 20 and 40).

The random phase approximation, or RPA, \cite{ring,RoweRMP,Marshalek} is an
appealing middle ground because it adds correlations to the
mean-field state and is numerically more tractable than full
diagonalization in a SM basis; it also exactly separates out
spurious modes induced by symmetry breaking. Because RPA implicitly
treats 2-particle, 2-hole correlations in the ground state, one might imagine RPA
to be an approximation to a more sophisticated microscopic model,
a kind of cleverly truncated interacting shell model so to speak.
Unfortunately, this image is erroneous because, unlike HF and the
interacting SM, RPA is not a variational theory, and in fact,
RPA can overestimate the `exact' binding energy. (This of course
supposes one has the `correct' effective interaction in the first
place, a big question we will not attempt to answer).

Surprisingly, tests of HF+RPA calculations of binding energies in
the literature have been scant and all have been against simple
models: the Lipkin-Meshkov-Glick (LMG) model 
\cite{LipkinNPA, RowePR1968b}, a schematic interaction in a 
small shell-model space \cite{RowePR1969} (here HF+RPA did 
poorly in describing the ``exact'' binding energy), a generalized 3-level LMG 
model   \cite{bertschPRC2000} and a
two-level pairing model \cite{bertschNPA2000}.  
Based particularly on the success of the latter two cases, Bertsch and 
Hagino advanced the idea of using the
mean-field theory corrected through RPA for nuclear binding-energy 
systematics \cite{bertsch2000}.

 This history motivates us to test the
accuracy of RPA with respect to other microscopic calculations. We
present a thorough study, comparing RPA to exact SM calculations
in a large number of non-trivial systems.


The paper is organized as follows. Section II presents the model space
and the interaction used, shortly reviews the approach and compares
the RPA and exact correlation energies for several nuclei.
We find that RPA reproduces
well the ground state energy for a large number of nuclides. 
Because the HF+RPA occasionally misses the binding energy, we attempt in 
Section III to determine if the accuracy of the HF+RPA 
binding energy correlates with properties of the HF state, 
such as deformation geometry or the overlap of the HF state
with the exact ground state. We find no such relationship. 
And because RPA is often said to ``restore'' broken symmetries, 
in section IV we compare the kinetic energy of the
RPA zero modes (associated with symmetry restoration) with the 
``kinematic'' correlation energy
obtained by projecting the HF state onto good $J$. 
Finally, we summarize our conclusions in section V.

\section{correlation energy in rpa}

In the interacting shell model one restricts the model space
to a limited set of valence fermion states outside an inert core, 
interacting via a residual two-body Hamiltonian.  We consider nuclei 
throughout the $sd$-shell, with an $^{16}$O core and using the 
Wildenthal USD interaction \cite{wildenthal}, and in the 
lower $pf$-shell, with a $^{40}$Ca core and using the 
modified KB3 interaction \cite{KB3}.
For each nucleus
investigated in this paper we are able to perform a full 0$\hbar\omega$
diagonalization; in the same valence space we calculate the HF solution
using separate Slater determinants for protons and neutrons.
In this restricted model there is no mixing of radial degrees of freedom,
the only degrees of freedom being angular. We emphasize  that
this is nonetheless a much more sophisticated testing ground 
for HF+RPA than the usual toy models.

Several textbooks  discuss extensively the matrix formulation
of RPA (see e.g., \cite{ring}),
so we  only summarize.
One starts with a self-consistent mean-field solution,
a Hartree-Fock Slater determinant, which
can break symmetries of the Hamiltonian and which ignores correlations.
The \textit{correlation energy} is the additional binding energy
that arises from particle-hole correlations. For example, if
one performs a shell-model diagonalization in the full space,
then the shell-model correlation energy, which we consider to be
the exact correlation energy, is
\begin{equation}
E^{corr}_{SM}=E_{HF}-E_{SM}.
\end{equation}

To derive the correlation energy in RPA, one treats the energy surface
around the Hartree-Fock minimum
as a multi-dimensional harmonic oscillator, which allows one to
compute excited states as 1-particle, 1-hole (1p-1h)
configurations, as well as to
compute corrections to the ground state energy. This last is approximately the 
correlation energy from an admixture of 2p-2h configurations. 
Because in the mapping to a multi-dimensional 
harmonic oscillator one has bosonized the energy surface, RPA does 
not rigorously enforce the Pauli principle, so RPA implicitly assumes the 
admixture of 2p-2h states into the ground state is small. 


The resulting RPA matrix equation is well-known:
\begin{equation}
\left(
\begin{array}{cc}
A & B \\
-B^* & -A^*
\end{array}\right)
\left(
\begin{array}{c}
X\\
Y
\end{array}\right)=
\Omega
\left(
\begin{array}{c}
X\\
Y
\end{array}\right),
\label{RPAmatrix}
\end{equation}
with 
\begin{eqnarray}
A_{nj,mi} \equiv \left \langle HF  \left |
\left [ \hat{c}^\dagger_j \hat{c}_n , [\hat{H} , \hat{c}^\dagger_m
\hat{c}_i ] \right ]\right | HF  \right \rangle, \nonumber \\
B_{nj,mi} \equiv \left \langle HF  \left |
\left [   [\hat{H} , \hat{c}^\dagger_n
\hat{c}_j ], \hat{c}^\dagger_m \hat{c}_i \right ]
\right | HF  \right \rangle.
\label{ABdefn}
\end{eqnarray}

Most applications of RPA are to transition strengths and 
excitation energies, but one can also derive 
the RPA correlation energy: 
\begin{equation}
E^{corr}_{RPA}=\frac{1}{2}{\rm Tr}(A) - \frac{1}{2}\sum_{i}\Omega_i.
\label{ecorr1}
\end{equation}
One can also compute the correlation energy in the response-function 
formulation of RPA \cite{ShimizuPRL2000}.

A particular feature of the RPA is the appearance of zero frequencies
which correspond to conserved quantities, sometimes interpreted as
restoration of broken symmetries \cite{RoweRMP,Marshalek}.
In our restricted model space, the mean field can break
the rotational symmetry; therefore, the number of zero
RPA frequencies is related to the shape of the HF solution. Thus,
when the HF solution is spherical, there are no zero modes;
when axially symmetric, we get two zero eigenfrequencies, while
for triaxial shapes we obtain three. The zero modes require careful
treatment, and one can rewrite Eqn.~(\ref{ecorr1}): 
\begin{equation}
E^{corr}_{RPA}=\sum_{\nu(\Omega_\nu\neq0)}\sum_{mi}\Omega_\nu |Y_{mi}^\nu|^2 +
\sum_{\mu(\Omega_\mu=0)}\frac{\langle {\cal P}_\mu^2\rangle}{2M_\mu},
\label{ecorr2}
\end{equation}
where ${\cal P}_\mu$ are the canonical momentum operators 
corresponding to zero modes
(for details, see  \cite{ring}).
Eqns. (\ref{ecorr1}) and (\ref{ecorr2}) are equivalent, 
but the latter explicitly segregates the 
contribution from zero modes, that is, from restored symmetries
\cite{bertschPRC2000,bertschNPA2000,ring,RoweRMP,Marshalek}.
We compare the kinetic energy due to the zero modes against the
correlation energy obtained
through projection of the HF state onto good $J$ in section IV.

Ideally one would hope for a good agreement between $E^{corr}_{RPA}$
and $E^{corr}_{SM}$. 
Figures 1-5 compare the exact and RPA correlation energies for nuclides
in the $sd$- and $pf$-shells respectively. Note that it is not our intention
at this moment to test against the experimental data.
In the $sd$-shell all the nuclei are numerically
tractable for a full diagonalization;
for an easier comparison we grouped into even-even (Fig. 1),
even-odd (Fig. 2), odd-odd (Fig. 3) and oxygen isotopes, that is, 
only neutrons active (Fig. 4).
In the $pf$-shell the dimension
of the SM basis rapidly increases with increasing the number of particles
in the valence shell, limiting the number of exact cases 
we can hand; in Fig. 5(a) we present nuclides with both
protons and neutrons in the active $pf$ space, while Fig. 5(b) contains
calcium isotopes (neutrons only).

\begin{figure}
\centering
\includegraphics[scale=0.95]{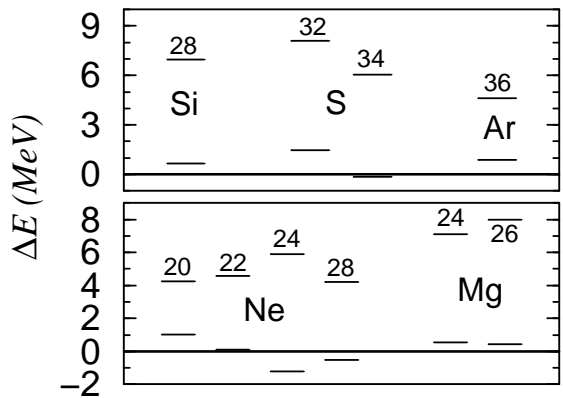}
\caption{HF and HF+RPA energies relative to the 
exact SM ground-state energy, for even-even nuclides 
in the $sd$ shell. The exact SM g.s. energy is set at zero, 
the upper bars are the HF energy and the lower bars the HF+RPA energy.}
\end{figure}

\begin{figure}
\centering
\includegraphics[scale=0.95]{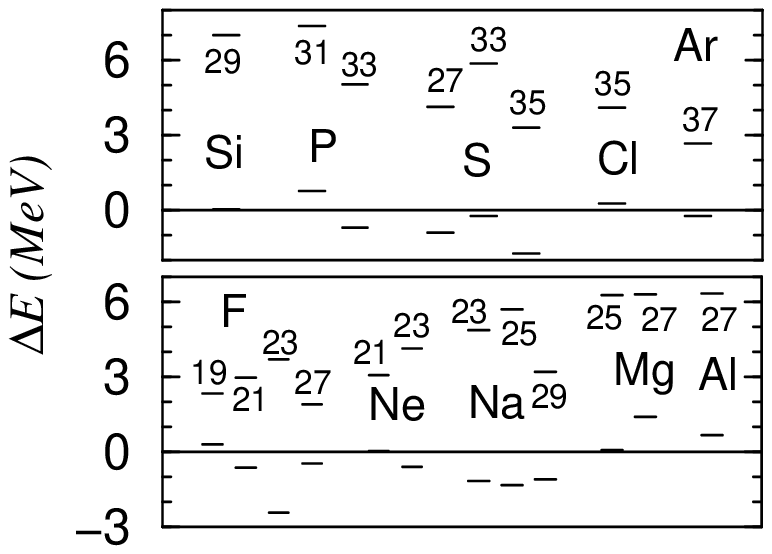}
\caption{
Same as in Fig. 1 for odd-A nuclides in the $sd$ shell.
}
\end{figure}

\begin{figure}
\centering
\includegraphics[scale=0.95]{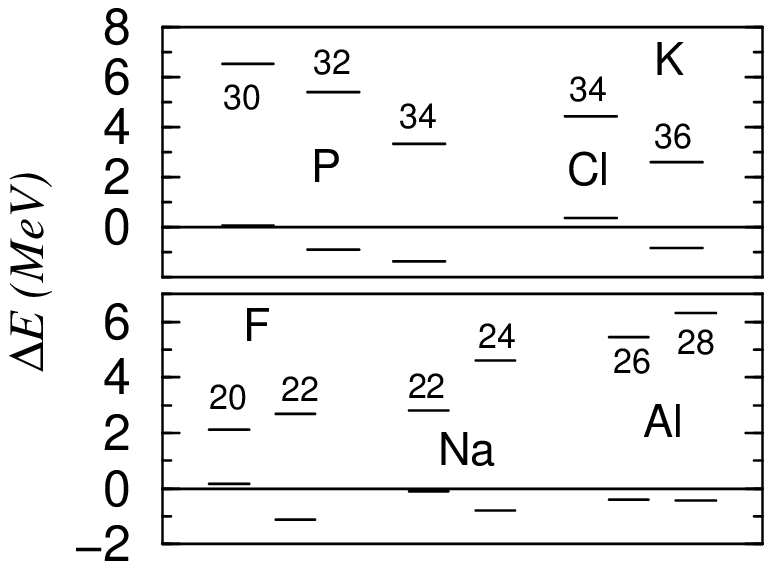}
\caption{
Same as in Fig. 1 for odd-odd nuclides in the $sd$ shell.
}
\end{figure}

\begin{figure}
\centering
\includegraphics[scale=0.95]{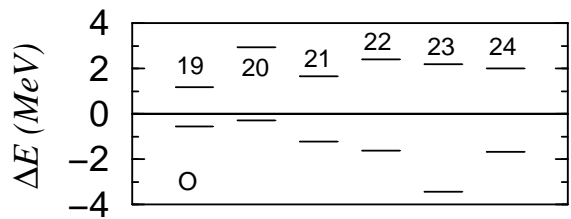}
\caption{
Same as in Fig. 1 for oxygen isotopes (that is, 
neutrons only). 
}
\end{figure}

\begin{figure}
\centering
\includegraphics[scale=0.95]{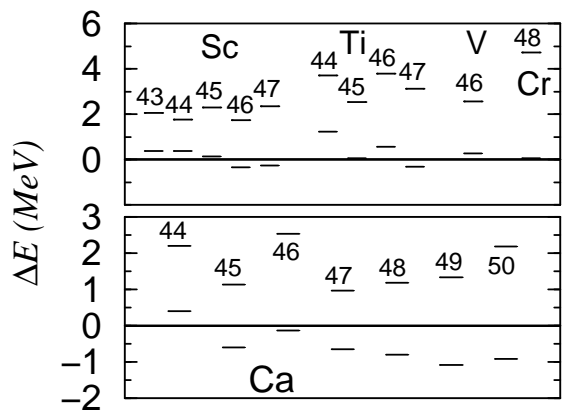}
\caption{
Same as in Fig. 1 in the $pf$-shell for 
(a) nuclides with both protons and neutrons in the valence space 
and (b) calcium isotopes (neutrons only). 
}
\end{figure}

The general trend is a good agreement between $E^{corr}_{SM}$
and $E^{corr}_{RPA}$; in each group, except for oxygen,  there is only one
or two nuclides where RPA significantly overestimates or underestimates the
correlation energy. In the $sd$-shell, if one excludes the oxygen results,
the rms deviation for 41 nuclides is 870 keV; in the lower $pf$-shell,
excluding calcium, the rms deviation is 480 keV for 11 nuclides.
There was not a significant
difference between even-even, odd-odd, and odd-even/even-odd nuclides.
The single-species results were significantly worse, however: 1800 keV for
6 oxygen isotopes and 730 keV for 7 calcium isotopes. We search for
potential explanations for this difference, as described in section III
below, but found none.

Despite these limitations, based on the results presented in Figs. 1-5,
one can assert that RPA seems to be a promising
candidate for a global microscopic theory of the ground state energies,
as advanced in \cite{bertsch2000}.

\section{Analysis of RPA accuracy}

Despite our overall good agreement, RPA sometimes fails to accurately 
reproduce the exact correlation energy. In this section we search 
for measures that correlate with the success or failure of 
RPA, which may in turn point towards possible solutions.  For example, 
if RPA fails to describe pairing correlations, one may turn to 
Hartree-Fock-Bogoliubov + quasi-particle RPA.  For another example, one 
might expect RPA, which implicitly assumes only small 2p-2h correlations, to 
be more accurate when the original HF state has a large overlap with the 
exact ground state wavefunction.  
If this were the case, one 
might correct through some self-consistent or renormalized RPA scheme. 

To quantify our search use the ratio
\begin{equation}
R=\frac{E^{corr}_{RPA}}{E^{corr}_{SM}}.
\end{equation}
$R=1$ when HF+RPA agrees with the exact shell-model binding energy.

We begin by considering how well RPA works with different geometries 
of the mean-field solution. In particular, how well does the RPA 
correlation energy account for (``restore'') broken symmetries?
Table I lists the deformation geometry  
(spherical, prolate, oblate, triaxial) for various 
nuclides; we compute $\beta$, $\gamma$ by diagonalizing the 
mass quadrupole tensor of valence nucleons for the HF state \cite{deform_method}. 
$R$ is close to
one for a wide range of deformation parameters.  In fact, 
if anything RPA is \textit{less} reliable for a spherical mean-field  
geometry, overestimating the  correlation energy 
by as much as a factor of two. 
This result is surprising. In exact shell-model calculations, the 
excited states for these ``spherical'' nuclei  
are dominated by 1-particle 1-hole configurations, 
so one might expect RPA to be more successful than for ``deformed'' 
nuclei.

\begin{table}[h]
\caption{Correlation energy ratio for select nuclides in $sd$ and
$pf$ shells with different geometries of the HF solution.}
\begin{ruledtabular}
\begin{tabular}{cccc}
Nucleus & $\beta$ & $\gamma$ (degrees) & $\frac{E_{HF}-E_{RPA}}{E_{HF}-E_{SM}}$
\\ \hline
$^{22}$O  & 0 & -- & 1.67 \\
$^{24}$O  & 0 & -- & 1.83 \\
$^{20}$Ne & 0.46 & 0 & 0.75 \\
$^{22}$Ne & 0.33 & 0 & 0.97 \\
$^{23}$F  & 0.11 & 60 & 1.65 \\
$^{24}$Na & 0.24 & 13 & 1.17 \\
$^{24}$Mg & 0.29 & 14 & 0.92 \\
$^{26}$Al & 0.20 & 33 & 1.07 \\
$^{28}$Si & 0.20 & 60 & 0.90 \\
$^{31}$P  & 0.14 & 40 & 0.81 \\
$^{32}$S  & 0.13 & 32 & 0.82 \\
$^{34}$S  & 0.09 & 48 & 1.02 \\
$^{34}$Cl & 0.10 & 42 & 0.91 \\
$^{36}$Ar & 0.09 & 60 & 0.81 \\
$^{43}$Sc & 0.28 & 60 & 0.80 \\
$^{44}$Ti & 0.44 & 0 & 0.66 \\
$^{45}$Ti & 0.38 & 0 & 0.97 \\
$^{46}$Ti & 0.35 & 0 & 0.85 \\
$^{46}$Ca & 0.12 & 0 & 1.04 \\
$^{48}$Ca & 0 & -- & 1.66 \\
\end{tabular}
\end{ruledtabular}
\end{table}

All of the spherical nuclei we found in the $sd$-shell were oxygen isotopes
(valence neutrons only); this is not surprising as the proton-neutron
interaction is well-known to induce deformation.
To pursue this issue further, we took $^{28}$Si, which normally 
has an oblate mean-field solution, 
and lowered the $0d_{5/2}$ single-particle energy, until
the mean-field solution became spherical (filled $d_{5/2}$ for both protons and
neutrons). In Fig. 6 we plot $R$ as function of the $d_{5/2}$ single
particle energy.  Once again $R$ is closer to 1 for the deformed regime
than for the spherical, or closed-shell, regime.

\begin{figure}
\centering
\includegraphics*[scale=0.7]{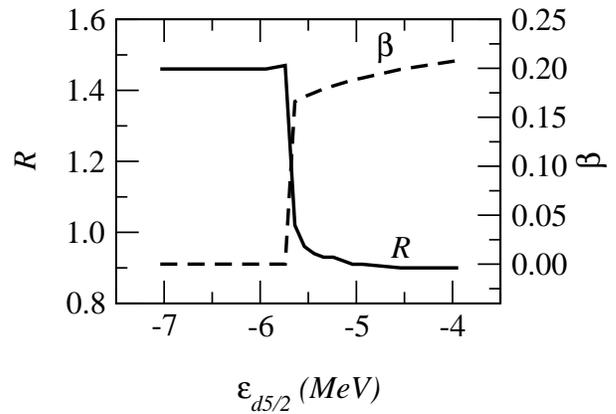}
\caption{Phase transition in $^{28}$Si as the $d_{5/2}$ single-particle
energy is lowered relative to the other single-particle energies.
The solid line is $R$, the ratio of  correlation energies,
while the dashed line is the deformation $\beta$.
}
\end{figure}

Now, as mentioned at the beginning of this section,
one possible explanation for the poor results for spherical
nuclides is inadequate treatment of pairing.
Because the onset of pairing is non-perturbative one might not expect
RPA to describe pairing correlations well. In order to test this idea,
we computed the expectation value of the pairing
Hamiltonian  in both the exact shell model calculation and in HF+RPA
\cite{JohnsonTBP}. We found, surprisingly, \textit{better} agreement for
oxygen and calcium nuclides and considerably poorer results for nuclides with
more deformation and better RPA correlations energies. Hence we cannot
point conclusively to pairing as the culprit.

Another place RPA may stumble is in its implicit assumption that 
the admixture of 2p-2h correlations in the ground state is small. 
This shows up in two places, in the bosonization of the energy surface 
about the Hartree-Fock minimum, which does not enforce the Pauli principle, 
and, in the computation 
of the RPA \textbf{A} and \textbf{B} matrices, 
the replacement of the RPA wavefunction by the HF state 
in equation (\ref{ABdefn}).  
To test this possibility, we calculate the overlap  between the HF state
and the exact wave function. This cannot be done by simply taking the projection
\begin{equation}
{\cal O}=\langle HF | SM \rangle.
\end{equation}
In our SM calculations, we took the basis states to have 
$J_z=0$ for $A$ even and $J_z=1/2$ for $A$ odd; 
the HF state is a superposition of
states with good $J_z$ which are not restricted to the SM values.
Therefore, any rotation of the Slater determinant can change the
weight of $J_z$ components so that ${\cal O}$ is not invariant,
although both $E^{corr}_{RPA}$ and $E^{corr}_{SM}$ are. We
therefore define the overlap as
\begin{equation}
O={\rm Max}\left(\langle HF | SM \rangle\right),
\label{Overlap}
\end{equation}
where the maximum is taken when rotating about the $y$ axis. Note
that when the exact ground state has $J=0$, ${\cal O}$ is independent 
of orientation. 

\begin{figure}
\centering
\includegraphics[scale=0.80]{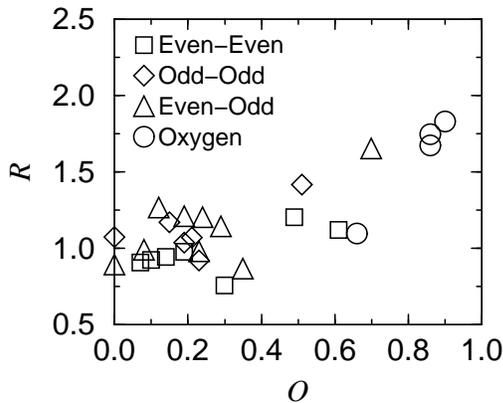}
\caption{
The correlation energy ratio $R$ vs. the overlap $O$, defined
in Eq.~(\ref{Overlap}),  for select nuclei in $sd$ shell.
}
\end{figure}

Fig. 7 shows $R$ as function of the overlap $O$; here we have
included only nuclides in the $sd$-shell, but the results for $pf$ are similar.
$O$ varies from close to zero to close to one, but no compelling
correlation with $R$ appears. That is to say, the accuracy of the
RPA correlation energy does not seem predicated on how well the HF state
approximates the exact ground state wavefunction.

A similar measure would be
the amount of correlations in the RPA wavefunction, that is,
$N_{ph}^{RPA} =\sum_\nu\sum_i |Y_{mi}^\nu|^2$.  In principle, RPA is more accurate when
$N_{ph}^{RPA}$ is small, but, as in Fig.~7, we find no correlation between
$N_{ph}^{RPA}$ and $R$. Because of this, it is not not obvious that any
renormalized, self-consistent, extended, etc., RPA scheme would in practice
yield substantial improvements.

\section{Kinetic energy of spurious modes}

One of the features that makes RPA appealing for correcting the mean-field
results is its capacity to identify and separate out,
as zero excitation energy, the modes associated with conserved quantities.
Consequently, one can identify in Eq.~(\ref{ecorr2}) a
term in the correlation energy corresponding to the kinetic energy of the
spurious solution
\begin{equation}
KE_0=\sum_{\mu(\Omega_\mu=0)}\frac{\langle {\cal P}_\mu^2\rangle}{2M_\mu}.
\label{KEspurious}
\end{equation}

On the other hand,
the simplest way to correct the broken symmetries in the mean-field solution
is to project it onto good quantum numbers. Table II presents the correlation
energy obtained by projection of the HF Slater determinants onto good $J$,
$E^{corr}_{proj}$, and the kinetic energy of the spurious modes given by
Eq. (\ref{KEspurious}); for comparison we have also included the exact
correlation energy missing from the mean-field state.  We conclude
that $KE_0$ cannot be identified with the correlation energy
obtained by direct restoration of symmetries obtained by projecting
the HF state onto good quantum numbers.  This may have relevance when if
one goes to a multi-$\hbar \omega$ space and wants to remove contamination
by spurious center-of-mass motion.

\begin{table}[h]
\caption{Kinetic energy of RPA zero modes compared with the correlation energy
obtained after projection of the HF state on good $J$.}
\begin{ruledtabular}
\begin{tabular}{cccc}
Nucleus & $E^{corr}_{proj}$ $(MeV)$ & $KE_0$ $(MeV)$ & $E^{corr}_{SM}$ $(MeV)$\\
\hline
$^{20}$Ne & 3.29 & 2.55 & 4.25 \\
$^{22}$Ne & 1.86 & 2.29 & 4.56 \\
$^{28}$Si & 2.67 & 3.85 & 6.97 \\
$^{24}$Mg & 5.34 & 4.65 & 7.10 \\
$^{32}$S  & 2.35 & 4.20 & 8.08 \\
$^{44}$Ti & 1.18 & 1.60 & 3.72 \\
$^{46}$Ti & 1.82 & 1.33 & 3.24
\end{tabular}
\end{ruledtabular}
\end{table}

\section{summary and outlook}

We have investigated the accuracy of HF+RPA for computing
the nuclear binding energies, by comparing against the exact
shell-model binding energies obtained through
full $0\hbar\omega$ diagonalization.  In simpler models
\cite{RowePR1968b,bertschPRC2000,bertschNPA2000}, the RPA
binding energy is generally, but not always \cite{RowePR1969},
satisfactory.  We found this also to
be true with our significantly more complicated shell-model Hamiltonians.

There were some non-trivial cases where RPA did not accurately estimate the
exact binding energy; however, aside from the rather unhelpful observation
that the inaccuracies tended to be worse for single-species cases,
we could find no useful correlate with the accuracy of RPA.  We did
find, however, that RPA tended to yield very good results for deformed
nuclides; despite this, however, the part of the RPA correlation
energy identifiable with rotational kinetic energy had no relation
with the energy gained from projection.

How could one make RPA more reliable? There are two
obvious steps.  The first is to attempt to build pairing directly
into the mean-field state through Hartree-Fock-Bogolyubov and
then computing the  quasiparticle-RPA correlations energy.
Because we found no relation between the accurate calculation
of the expectation value of the pairing Hamiltonian and the
accurate calculation of the total binding energy, however,
we are far from confident that HFB+QRPA will yield significant gains.

The other possibility is to correct RPA for its violation of the
Pauli principle as well as using the RPA ground state wavefunction
function in computing \textbf{A} and \textbf{B} matrices in 
equation (\ref{ABdefn}).
The literature documents numerous proposals along these lines:
self-consistent renormalization of the mean-field single-particle 
energies and
the residual interaction \cite{RoweRMP,scrpa};
treating the particle-hole amplitudes as variational
parameters \cite{Dreizler}; 
rederivation of the RPA equations via the ``number-operator method'' 
to account for ground state correlations 
\cite{KleinNPA1991};
and dressing the RPA phonons to account for ground state correlations 
\cite{DinhPRC2001}.
None of these methods have become widely accepted, and they overlap 
each other to a large degree, although the literature 
does not clarify similarities and differences. 
As in the case of ungarnished RPA, these proposals for 
improving RPA have only been tested against toys such as the 
Lipkin model \cite{LipkinNPA}. 
Perhaps by
implementing these various proposals for full-scale shell model
calculations one could compare their efficacy and practicality.
This we leave to future work.

The U.S.~Department of Energy supported this investigation through
grant DE-FG02-96ER40985.


\begin{thebibliography}{99}

\bibitem{MN95}  P.~M\"oller, J.R.~Nix, W.D.~Myers, and W.J.~Swiatecki, 
At.~Data Nucl.~Data Tables \textbf{59}, 185 (1995).

\bibitem{AP95} Y.~Aboussir, J.M.~Pearson, A.K.~Dutta, and F.~Tondeur, 
At.~Data Nucl.~Data Tables \textbf{61}, 127 (1995).

\bibitem{benchmark} 
H.~Kamada \textit{et al}., Phys.~Rev.~C \textbf{64}, 044001 (2001).

\bibitem{Aeq8nuclei} R.~B.~Wiringa, S.~C.~Pieper, J.~Carlson, and 
V.~R.~Pandharipande, Phys.~Rev.~C \textbf{62}, 014001 (2001).


\bibitem{Aeq12nuclei} P.~Navratil, J.P.~Vary, and B.R.~Barrett, 
Phys.~Rev.~Lett.~\textbf{84}, 5728 (2000); Phys.~Rev.~C \textbf{62},
054311 (2000).

\bibitem{HFmass} F.~Tondeur, S.~Goriely, J.M.~Pearson, and M.~Onsi, 
Phys.~Rev.~C \textbf{62}, 024308 (2000).

\bibitem{ring}
P. Ring and P. Shuck, \textit{The nuclear many-body problem}, 1st edition,
Springer-Verlag, New York 1980.

\bibitem{RoweRMP}
D.J. Rowe, Rev. Mod. Phys. \textbf{40}, 153 (1968).

\bibitem{Marshalek}
E.R. Marshalek and J. Weneser, Ann. Phys. \textbf{53}, 569 (1969).

\bibitem{RowePR1968b}
J.C. Parikh and D.J. Rowe, Phys. Rev. \textbf{175}, 1293 (1968).

\bibitem{LipkinNPA}
H.J. Lipkin N. Meshkov and A.J.Glick, Nucl. Phys. \textbf{62},
188 (1965).


\bibitem{RowePR1969}
N. Ullah and D.J.~Rowe, Phys. Rev. {\textbf 188}, 1640 (1969).

\bibitem{bertschPRC2000}
K. Hagino and G.F. Bertsch, Phys. Rev. C\textbf{61} 024307 (2000).

\bibitem{bertschNPA2000}
K. Hagino and G.F. Bertsch, Nucl. Phys. \textbf{A679} 163 (2000).

\bibitem{bertsch2000}
G.F. Bertsch and K. Hagino, Phys.~Atom.~Nucl.~\textbf{64},588 (2001) 588;
arXiv:nucl-th/0006032.



\bibitem{wildenthal}
B.H. Wildenthal, Prog. Part. Nucl. Phys. \textbf{11}, 5 (1984).

\bibitem{KB3} T.T.S.~Kuo and G.E.~Brown, Nucl.~Phys.~\textbf{A114}, 235 (1968); 
A.~Poves and A.P.~Zuker, Phys.~Rep.~\textbf{70}, 235 (1981).

\bibitem{ShimizuPRL2000}
Y.R. Shimizu, P. Donnati and R.A. Broglia, Phys. Rev. Lett. \textbf{85},
2260 (2000).

\bibitem{deform_method} W.E. Ormand, D.J. Dean, C.W. Johnson, G.H. Lang, 
and S.E. Koonin, 
Phys.~Rev.~C  \textbf{49}, 1422 (1994). 

\bibitem{JohnsonTBP}
C.W. Johnson and I. Stetcu, to be published.

\bibitem{scrpa}P.~Schuck and S.~Ethofer, Nucl.~Phys.~\textbf{A212}, 
269 (1973).



\bibitem{Dreizler}
R. M. Dreizler, A. Klein, F.R. Krejs and G.J. Dreiss,
Nucl. Phys. \textbf{A166}, 624 (1971).



\bibitem{KleinNPA1991}
A. Klein, N. R. Walet and G. Do Dang, Nucl. Phys. \textbf{A535},
1 (1991).

\bibitem{DinhPRC2001}
N. Dinh Dang and V. Zelevinsky, Phys. Rev. C \textbf{64}, 064319 (2001).





\end{thebibliography}
\end{document}